\documentclass{iaus}
\usepackage{graphicx}
\newcommand\tbce{T$_{\rm bce}$} 
\newcommand{\lsun}{\ensuremath{\, {L}_\odot}}
\newcommand{\mbol}{\ensuremath{\, {M}_{\rm bol}}}
\newcommand{\msun}{\ensuremath{\, {M}_\odot}}
\newcommand{\Msun}{\ensuremath{\, {M}_\odot}}

\newcommand{\ocen}{$\omega$~Cen}
\def\simgt{\lower.5ex\hbox{$\; \buildrel > \over \sim \;$}}
\def\simlt{\lower.5ex\hbox{$\; \buildrel < \over \sim \;$}}

\title[Lithium factories in the Galaxy] 
{Lithium factories in the Galaxy: novae and AGB stars}

\author[F. D'Antona \& P. Ventura]   
{Francesca D'Antona$^1$
 \and Paolo Ventura$^1$}

\affiliation{$^1$INAF -- Osservatorio di Roma, \\ via di Frascati 33,
I-00040 Monteporzio, Italy \\ email: {\tt dantona@oa-roma.inaf.it ventura@oa-roma.inaf.it}
}

\pubyear{2010}
\volume{268}  
\jname{Light elements in the Universe}
\editors{C. Charbonnel, M. Tosi, F. Primas \& C. Chiappini, eds.}
\begin{document}

\maketitle

\begin{abstract}
We review the state of the art in modelling lithium production, through the 
Cameron--Fowler mechanism, in two stellar sites: during nova explosions and in 
the envelopes of massive asymptotic giant branch (AGB) stars. We also show 
preliminary results concerning the computation of lithium yields from super--AGBs, and suggest that super--AGBs of metallicity close to solar may be the 
most important galactic lithium producers. Finally, we discuss how lithium 
abundances may help to understand the modalities of formation of the ``second 
generation" stars in globular clusters.

\end{abstract}

\firstsection 
\section{Introduction}

Although lithium is very fragile, its galactic abundance increases from $\log 
\epsilon$(Li)\footnote{ we use the notation $\log \epsilon$(Li)= $\log 
(N_{Li}/N_{H})+12.$} $\sim$2.2 at the surface of Population (Pop) II stars to $\log 
\epsilon$(Li)$\sim$3.3 or more in Pop I. Even if lithium is hidden in the 
atmospheres of Pop II, and its true primordial abundance is $\sim$2.7, a 
galactic production by $\sim$0.7dex is necessary.

The mechanism responsible for lithium production has been proposed by 
\cite[Cameron \& Fowler (1971)]{cameronfowler1971}: $^7$Be is produced by 
fusion of $^3$He with $^4$He, and rapidly transported to stellar regions where 
it can be converted into $^7$Li by k--capture. Notice, then, that the lithium 
production may last only until there is $^3$He\ available in the region of 
burning, and that the production ends when the $^3$He\ is all consumed. 

There are two main physical situations where this mechanism can produce enough 
lithium that it is important to investigate their role in the galactic 
production: the first one is the explosive hydrodynamical formation during the 
outbursts of novae (\cite[Arnould \& Norgaard 1975]{an1975}, \cite[Starrfield 
et al. 1978]{starrfield1978}), the second one is the hydrostatic, slow 
formation in the envelopes of asymptotic giant branch (AGB) stars, for which it 
was first proposed. In envelope models of AGB stars (\cite[Scalo et al. 
1975]{scalo1975}), in which the bottom of the convective envelope  reaches the 
hydrogen burning layers, and its temperature (\tbce) becomes as large as 
\tbce$\sim$40MK, the $^3$He($\alpha,\gamma)^7$Be chain acts. These models were 
able to explain the high lithium abundances found in some luminous red giants, 
and the process took the name of Hot Bottom Burning (HBB). 

In Section 2 we will resume the state of the art of the modelling of lithium 
production during nova outbursts, and in Section 3 we will deal with the AGB 
models, to understand whether they can account for the lithium galactic 
evolution. In addition, we will show new models of lithium production in super--AGB stars (\cite[Ventura \& D'Antona 2010]{vd2010}) and speculate on the 
possible role of these stars as efficient lithium factories. Finally, in 
Section 4 we will shortly summarize the problem of lithium in the ``second 
generation" stars of globular clusters. We will not consider here the 
different, slow mixing process also based on the \cite{cameronfowler1971} 
mechanism and named ``cool bottom burning" (e.g. \cite[Nollett et al. 
2003]{nollett2003}). This process can explain the lithium abundances seen in 
lower luminosity red giants (e.g. \cite[Wasserburg et al. 
1995]{wasserburg1995}, \cite[Sackmann \& Boothroyd 1999]{sackmann1999}), but 
its physical reasons are not well studied, while the nucleosynthesis in HBB is 
based on straightforward time--dependent mixing in standard convective regions.

\section{Nova outbursts}
\cite{schatzman1951} was the first to propose that the isotope $^3$He\ could 
play a role in nova explosion, in the context of a theory of novae powered by 
thermonuclear detonations. Arnould \& Noergaard (1975) proposed that the 
Cameron--Fowler mechanism, acting at the nova outburst, would produce a lithium 
abundance proportional to the  $^3$He\ abundance in the nova envelope. 
\cite{starrfield1978} showed that the mechanism could be efficient for outburst 
temperatures $>$150MK, and the fast ejection of the $^7$Be rich nova shell 
leads to $^7$Li production; they quantified the expected linear relation, 
between lithium and the $^3$He\ initial mass fraction  $X_{3i}$, as:
\begin{equation}
[Li/H] \simeq 200 \times X_{3i}/X_{3\odot} 
\label{eq1}
\end{equation}
where $X_{3\odot}$\ is the solar $^3$He\ mass fraction. \cite[D'Antona \& 
Matteucci (1991)]{dm1991} modelled the galactic evolution of lithium, including 
the contribution of novae according to this result. The argument below their 
modelization was very simple: the nova explosion occurs when a critical 
hydrogen rich envelope is reached on the white dwarf component of the nova 
binary, by accretion from its low mass companion. By losing mass, low mass 
stars expose the stellar regions in which the hydrogen burning p--p chain is 
incomplete, and thus bring to the surface the $^3$He\ accumulated in the 
envelope during the period preceding the mass transfer phase, and during the 
(slow) mass transfer phase itself (e.g. \cite[D'Antona \& Mazzitelli 
1982]{dm1982}). Thus \cite{dm1991} linked the lithium abundance produced in the 
outburst to the``delay time" between the formation of the white dwarf and the 
occurrence of nova outbursts. As a result, mainly the novae containing an 
``old" white dwarf, and therefore an old and $^3$He--rich low mass companion 
contribute to the galactic production of lithium, in agreement with the Li vs. 
[Fe/H] galactic relation.

Motivated by the \cite{dm1991} paper, \cite[Boffin et al. (1993)]{boffin1993} 
revisited the influence of $^3$He\ on the nova outbursts with simple one--zone 
models, but found out that equation \ref{eq1} was a large overestimate, due to 
two main reasons: 1) the neglect of the reaction $^8B(p,\gamma)^9C$\ in the 
\cite{starrfield1978} network, and 2) the increasing influence of the 
competitive reaction $^3He(^3He,2p)^4He$\ when $^3$He\ is enhanced in the nova 
envelope. Consequently, they found a milder dependence on the lithium 
production on the $^3$He:
\begin{equation}
{X(^7Li) \over X_0(^7Li)}  \simeq  1 + 1.5 \log {X(^3He) \over X_\odot(^3He)} 
\label{eq2}
\end{equation}
where $X$\ represents mass fractions, and $X_0(^7Li)$\ is the lithium 
production when the solar $^3$He\ abundance is adopted. Afterwards, 
\cite{hernanz1996} and \cite{jh1998} re-examined the problem with an implicit 
hydro-code including a full reaction network, able to treat both the 
hydrostatic accretion phase and the explosion stage. They considered both the 
case of white dwarfs having a carbon--oxygen core and the case of oxygen-- neon 
cores, showing that C--O cores are more efficient in the lithium production, as 
they have a shorter accretion phase, so that $^3$He\ is not destroyed 
efficiently, and more $^7$Be is produced. Overproduction of lithium is found, 
but its dependence on the initial $^3$He\ abundance still follows 
\cite{boffin1993} prescription. Including this revised lithium production in 
the galactic chemical evolution model, novae appear to be a modest lithium 
producer (\cite[Romano et al. 1999]{romano1999}).

\section{Luminous AGB stars}
Above a luminosity of $\sim 2 \times 10^4$\lsun, the bottom of the convective 
envelope during the AGB evolution reaches the H--shell burning region, and the 
nuclear reaction products are transported to the surface by convection. This is 
the perfect site of lithium production through the Cameron--Fowler mechanism 
(\cite[Iben 1973]{iben1973}, \cite[Sackmann et al. 1974]{sackmann1974}). While 
we have seen that \tbce$\sim$40MK is sufficient to produce lithium by HBB, if 
\tbce\ becomes larger, other important reactions take place.

About 65MK are necessary to convert carbon to nitrogen. The very luminous 
(\mbol$<-6$, that is L$>2 \times 10^4$\lsun) lithium--rich giants of the 
Magellanic Clouds (\cite[Smith \& Lambert 1989]{smith1989}, 
\cite[1990]{smith1990}, \cite[Smith et al. 1995]{smith1995}) are indeed M--stars, and not carbon stars. Their carbon star features may have been lost by 
CN processing in HBB. Carbon stars in the Clouds, in fact, populate only the 
region at \mbol$>-6$. Lithium rich -- oxygen rich AGB stars embedded in thick 
circumstellar envelopes have also been discovered in a Galactic sample, in a 
survey by \cite{gh2007}, aimed at obtaining spectroscopy of very massive AGB 
candidates.

A third possible processing occurs at even larger \tbce ($>$80MK), where H 
burns through the full CNO cycle. These very high temperatures are reached in 
low metallicity massive AGBs, and are possibly at the basis of the self--
enrichment process in globular clusters (\cite[Ventura et al. 
2001]{ventura2001}). The oxygen abundance in the envelopes of these AGB stars, 
and consequently in the matter ejected by wind or planetary nebula, is reduced, 
as we see in the ``anomalous" stars of galactic globular clusters, see Sect.4.
 
\begin{figure*}[b]
\begin{center}
\includegraphics[width=2.5in]{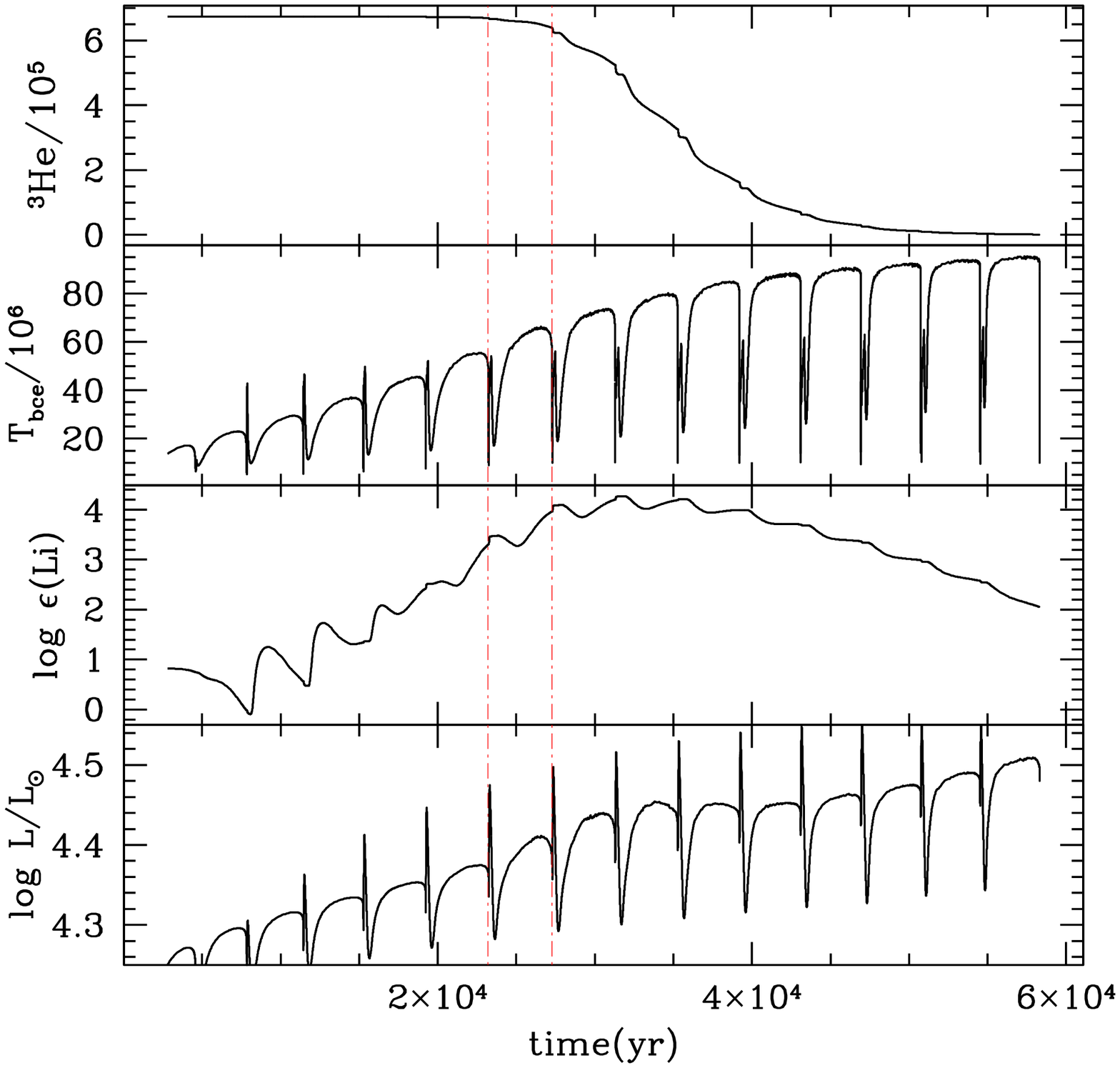} 
\includegraphics[width=2.5in]{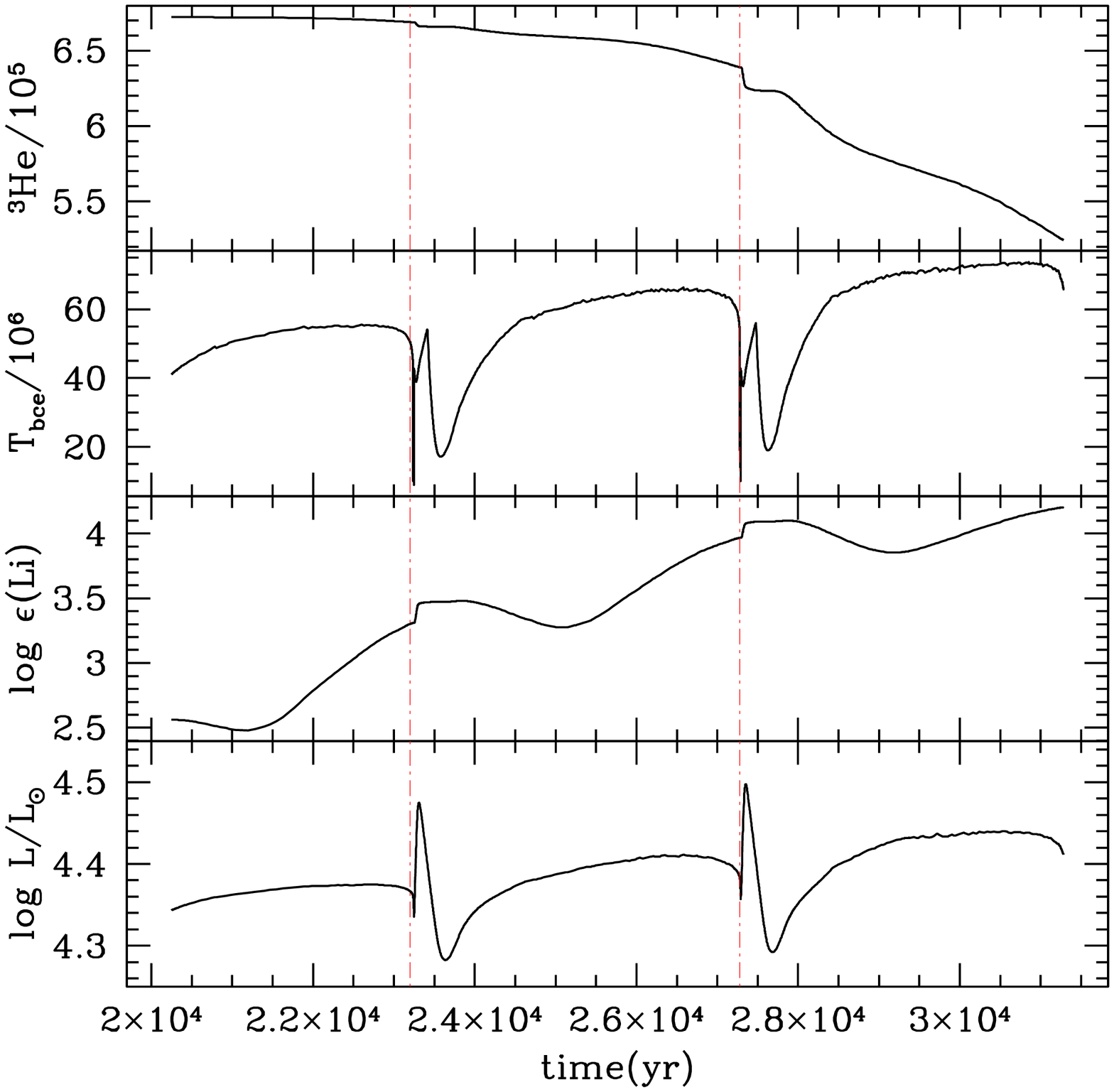} 
\caption{
From bottom to top panel we plot the luminosity, surface lithium, HBB 
temperature and $^3$He\ surface content along the AGB evolution of a star of 
5\Msun, metallicity Z=10$^{-3}$. The total duration of the phase of the most 
important lithium production lasts $\sim 20 \times 10^3$yr. The two vertical 
lines delimit the time interval displayed in the right panel. } 
\label{f1} 
\end{center} \end{figure*}

Modelling of lithium rich AGB stars first of all requires to treat non--
instantaneous mixing in the envelope, coupling the nuclear reaction network 
with the mixing process. This can be easily done by treating mixing as a 
diffusion. In Figure \ref{f1} we show the total phase of lithium production in 
a 5\msun\ star of metallicity Z=10$^{-3}$ (left side), and a zoom of the same 
figure between two thermal pulses (right side). We see that, when \tbce\ 
decreases, due to the ignition of the thermal pulse and the expansion of the 
envelope, the lithium abundance decreases. We can appreciate the delay time 
between the physical conditions in the burning region and the surface lithium, 
due to the non instantaneous mixing. The total phase of lithium production 
lasts more than 50$\times 10^3$yr, but the phase in which $\log 
\epsilon$(Li)$\simgt$3 lasts only $\sim 20\times 10^3$yr. Once the initial 
$^3$He\ present in the envelope is depleted, lithium production is over.

\begin{figure}[t]
\begin{center}
 \includegraphics[width=3.4in]{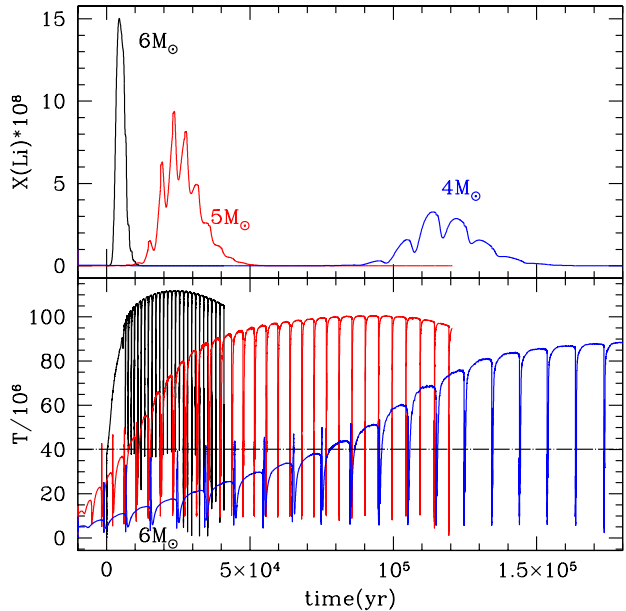} 
\caption{Temperature at the bottom of the convective layer (bottom) and lithium surface abundance
(top) as a function of the time for the masses 6, 5 and 4\msun, Z=10$^{-3}$, 
from left to right. Time is computed
from the beginning of the AGB phase, when the H--shell burning is reignited.
The horizontal line at T=40MK limits the temperature region for lithium production.}
   \label{f2}
\end{center}
\end{figure}

Lithium production and destruction depend on
\begin{enumerate}
\item the physical inputs, and mainly on the convection model: the higher is the
convection efficiency, the larger is \tbce\ and the larger is the efficiency of HBB
(see, e.g. \cite[Ventura \& D'Antona 2005]{ventura2005});
\item the initial mass (or, better, the initial core mass): it must be large enough to get HBB; 
\item the chemical inputs, mainly the metallicity and the envelope opacity. Fixed the mass,
the higher is the opacity (or the metallicity) the smaller is \tbce\, and the lower is the
efficiency of HBB.
\end{enumerate}

Figure \ref{f2} shows the mass dependence for a fixed chemical composition: the 
larger is the mass, the larger is \tbce\ and the stronger and faster is lithium 
production.
\begin{figure}[t]
\begin{center}
 \includegraphics[width=2.6in]{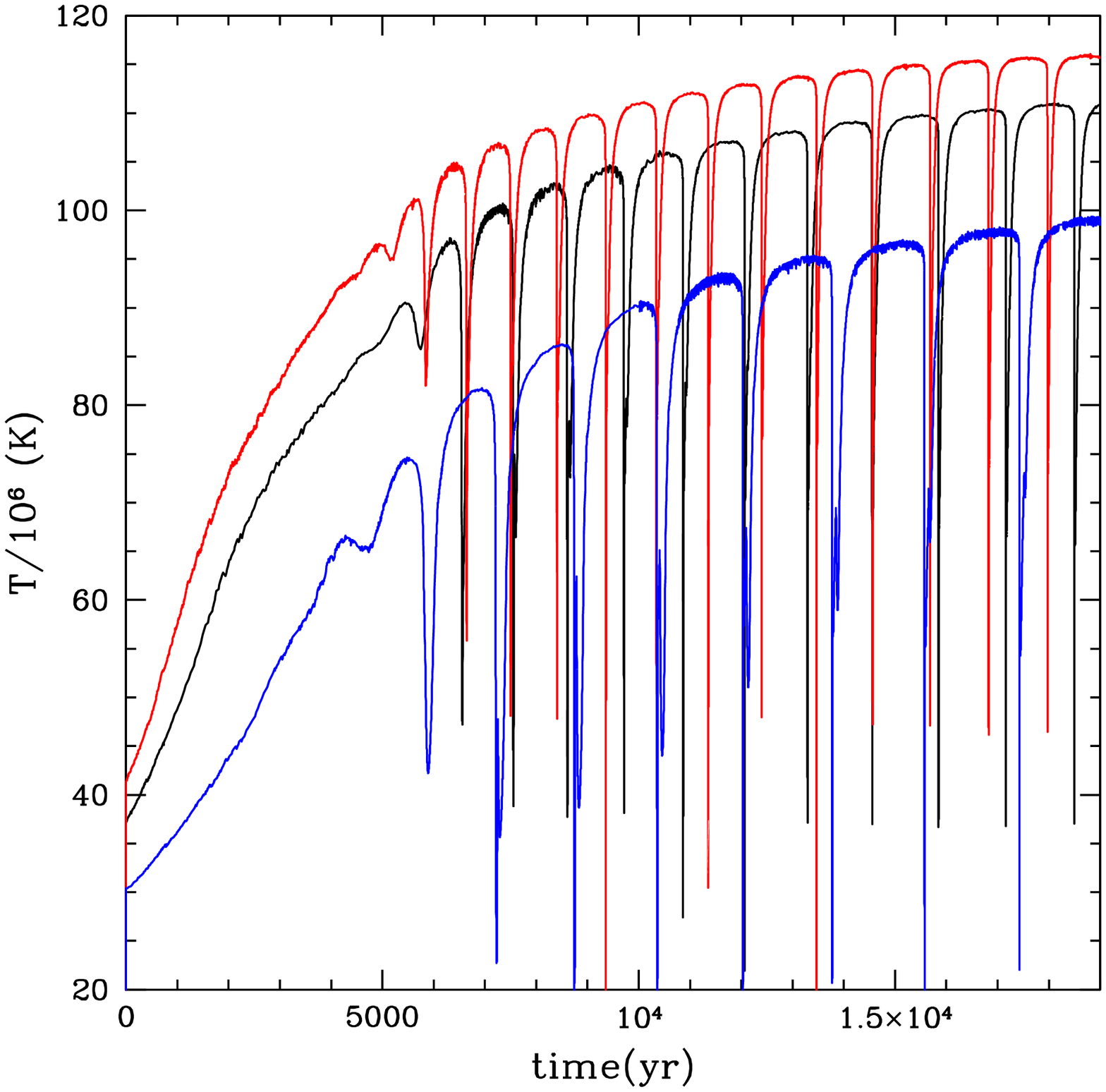} 
 \includegraphics[width=2.6in]{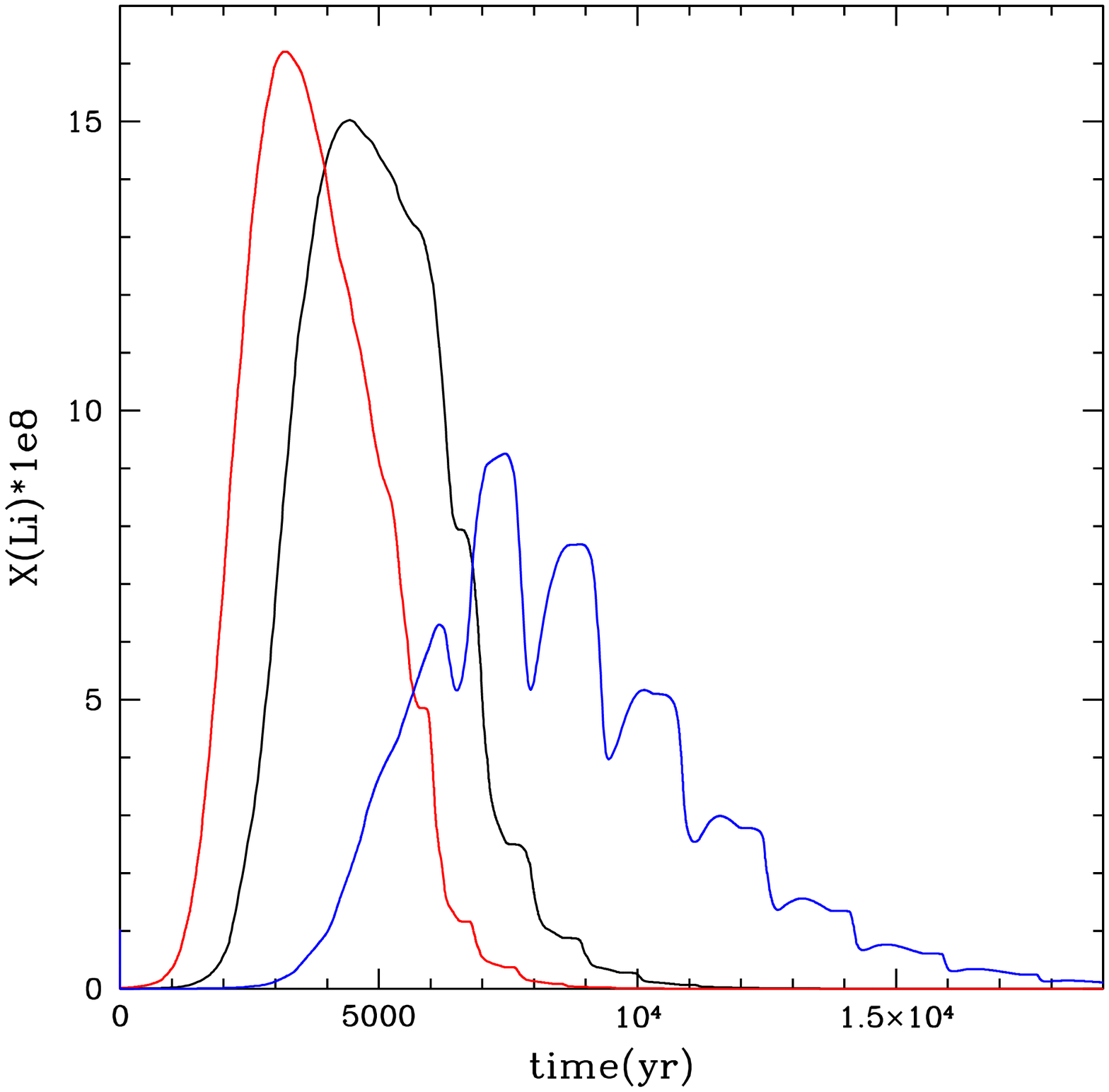} 
\caption{The evolution of 6\msun for metallicities Z=0.0006, Z=0.001 and Z=0.004 from top to
bottom is displayed. On the left side, we plot \tbce, on the right side the lithium mass fraction X(Li).
}
   \label{f3}
\end{center}
\end{figure}

Figure \ref{f3} shows the dependence on the metallicity, at fixed mass 
M=6\msun. Increasing the opacity (and Z), \tbce\ decreases, and the lithium 
production is lower but more extended in time.

\begin{figure*}[b]
\begin{center}
 \includegraphics[width=2.6in]{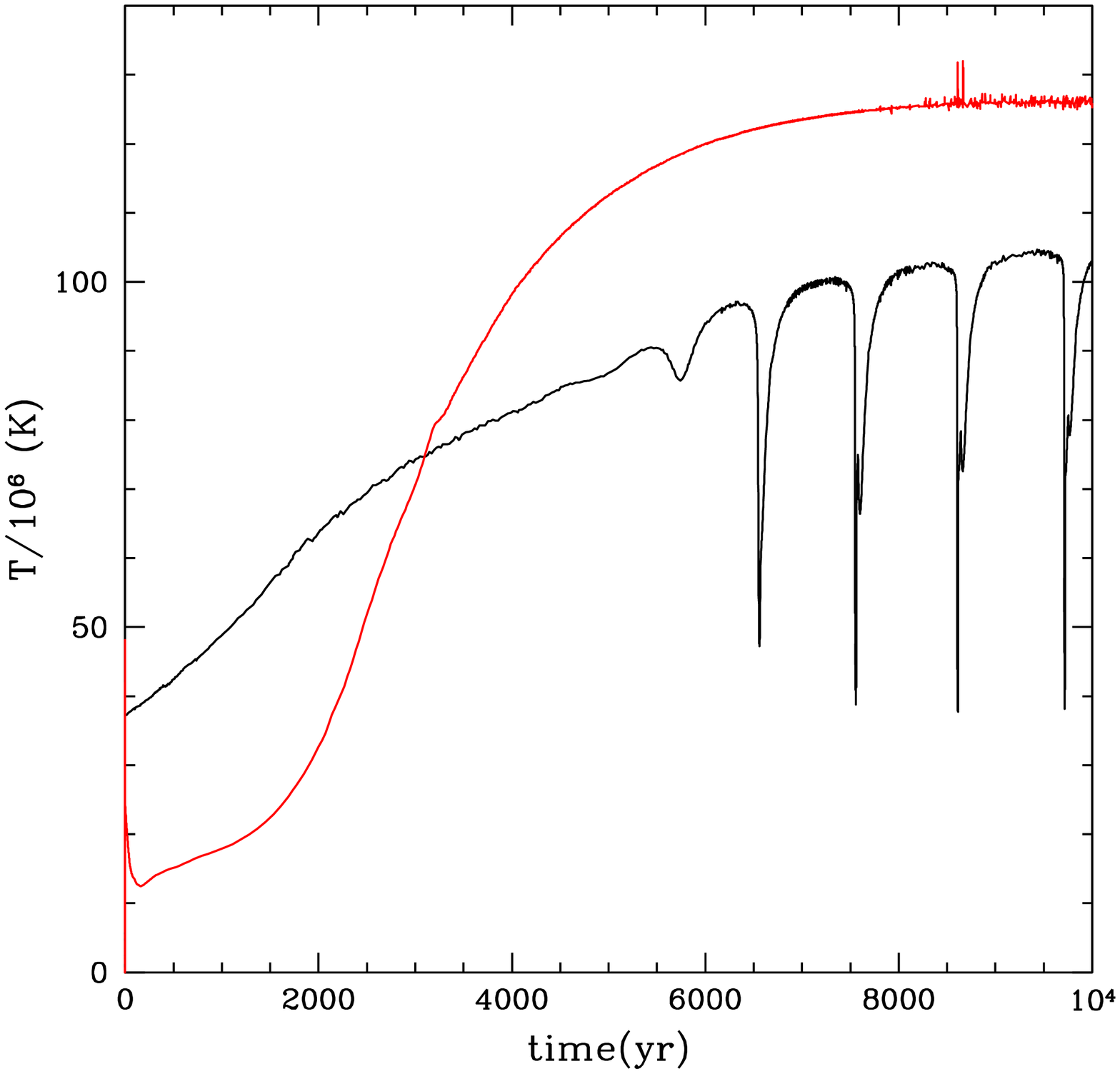} 
 \includegraphics[width=2.6in]{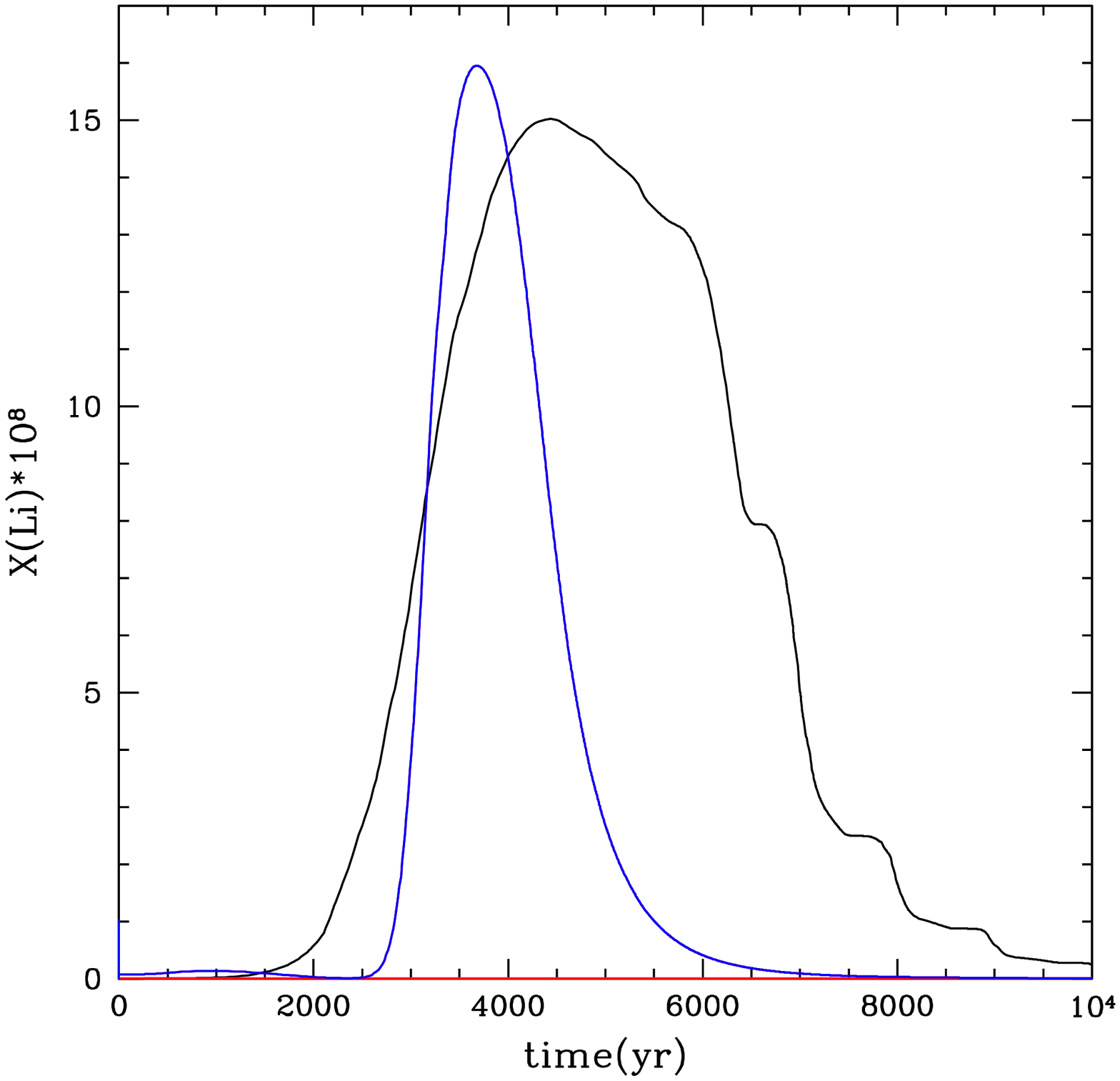}
 \caption{The same as Fig. 3 for Z=0.001 and masses 6\msun (lower curve) and 
 a super--AGB model of 7.5\msun.
 }
   \label{f4}
\end{center}
\end{figure*}

The computation of lithium production during the super--AGB evolution has been 
recently achieved by \cite[Ventura \& D'Antona (2010)]{vd2010} for Z=10$^{-3}$. 
The results are very interesting, as we see in Fig. \ref{f4} for a mass of 
7.5\msun. Lithium achieves very large abundances, due to the very high \tbce, 
and the Li--rich phase occurs even before the star begins the thermal pulse 
phase.

\begin{figure}
\begin{center}
\includegraphics[width=7cm]{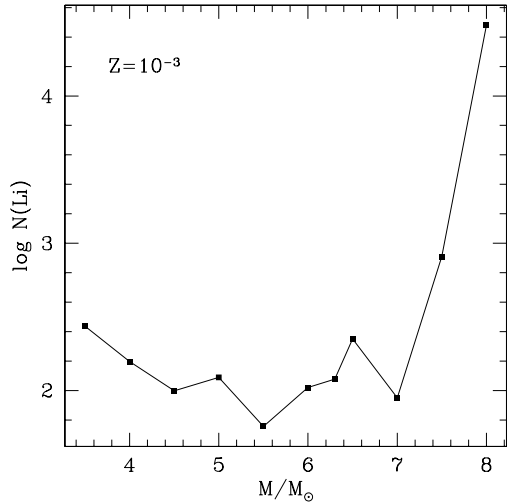}
\caption{Lithium abundance averaged on the ejected envelope mass as a function of the total
mass for Z=0.001}
      \label{f5}
\end{center}
\end{figure}

Of course `production' does not mean `yield': two ingredients are important: 
how much lithium is made, and how long it lasts, so that mass loss can recycle 
it into the interstellar medium. Consequently, the lithium yield is very 
dependent on the mass loss rate: larger rates during the phase of lithium 
production provide a higher lithium yield. Unfortunately, mass loss is another 
great uncertainty in the computation of stellar models. In Figure \ref{f6} we 
show as open (red) circles at [Fe/H]=--1.3 the average lithium abundance in the 
ejecta of models of 4, 5 and 6\msun\ with three different mass loss 
formulations: the middle points refer to the mass loss rate suggested by 
Bl\"ocker (1995), who extends Reimers' recipe to describe the steep increase of 
mass loss with luminosity as the stars ``climb" the AGB. The full expression, 
for Mira's periods exceeding 100d, is
\begin{equation}
\dot M=4.83 \times 10^{-22} \eta_R M^{-3.1}L^{3.7}R
\end{equation}
where $\eta_R$ is the free parameter entering the Reimers' (1977) prescription. 
In the ``standard" models of Fig. \ref{f6} we adopt $\eta_R=0.02$, according 
to a calibration based on the luminosity function of lithium rich stars in the 
Magellanic Clouds given in \cite[Ventura et al. (2000)]{ventura2000}. The 
highest points in the figure are obtained for the extreme value of 
$\eta_R=0.1$, while the models adopting the Vassiliadis \& Wood (1993) mass 
loss rate are the lowest ones. We see then that the absolute values of the 
lithium yields must be considered highly uncertain (see also \cite[Ventura et al. 2002]{ventura2002}). The global behaviour of the average lithium 
abundance in the ejecta, as a function of the initial mass is given in Fig. 
\ref{f5} for the models computed with the standard mass loss ($\eta_R$=0.02) 
prescription and Z=10$^{-3}$.
Increasing the mass, the average abundance first decreases, due to the faster 
consumption of $^3$He, in spite of the larger abundances reached in the phase 
of production. For the super--AGB masses, the average abundance increases, and 
may also become very large, both due to the stronger production and to the huge 
mass loss rate achieved by the largest core masses.

Based on these first computation of the super--AGB phase, we can make a 
prediction on the lithium galactic evolution: which are the best producers? The 
larger is the mass, the higher is lithium during the production phase. On the 
other hand, the shorter is the duration of the Li rich phase, the smaller is 
the lithium yield. Increasing the metallicity, \tbce\ becomes smaller, and the 
duration of the lithium production phase is longer. For large core mass (and 
thus high luminosity) the mass loss rates become larger and larger. So we 
should expect that the Li yield is positively correlated both with metallicity 
and core mass, and that the super--AGB stars of metallicity close to solar are 
possibly great producers. The possible consequences for galactic Li production 
are described in the talk by Francesca Matteucci in this book.
 
\section{Lithium and AGB stars in globular clusters}

Globular Clusters (GCs) so far examined show spectroscopic evidence for the 
presence of two stellar generations: a First Generation (FG) having ``normal" 
abundances, similar to those of halo stars of the same metallicity, and a 
Second Generation (SG) whose abundances are more spreaded, and bear the sign of 
hot CNO processing, with an often very significative oxygen reduction, evidence 
for the action of the Ne--Na cycle and sometimes of the Mg--Al cycle (see, e.g. 
\cite[Gratton et al. 2004]{gratton2004}). The SG contains at least 50\% of the 
cluster stars (Carretta et al. 2009a,b). At low metallicity, in the most 
massive AGB stars, \tbce\ becomes larger than $\sim$80MK, and the ON chain of 
the CNO cycle becomes active. In these envelopes, oxygen is cycled to nitrogen, 
and its abundance can be dramatically reduced. Thus some models for the 
formation of the different populations attribute the presence of ``anomalous" 
stars with low oxygen and high sodium, to a SG including matter processed by 
HBB (e.g. \cite[Ventura et al. 2001]{ventura2001}). Other models attribute the 
formation of the SG to the ejecta of fast rotating massive stars (FRMS, see 
e.g. \cite[Decressin et al. 2007a]{decressin2007}), or even to pollution from 
gas expelled during highly non conservative evolution of massive binaries 
(\cite[De Mink et al. 2009]{demink2009}), although this latter model in 
particular can not explain the very high fraction of SG stars present in most 
of the GCs so far examined.

The lithium yield from AGB stars of different mass may contribute to understand 
the role (if any) of these stars in the formation of the SG in GCs. It is 
commonly believed that the polluting matter must be diluted with pristine 
matter to explain the abundance patterns, such as the Na--O anticorrelation 
(Prantzos \& Carbonnel 2006, D'Antona \& Ventura 2007). If the progenitors of 
the SG stars are massive stars, they have destroyed their original lithium, and 
the lithium in the SG must be due to the mixing with pristine gas. If instead 
the progenitors are massive AGB stars, they may have a non negligible lithium 
yield, that must be taken into account in the explanation of the SG abundances.

\begin{figure}
\begin{center}
\includegraphics[width=3.8in]{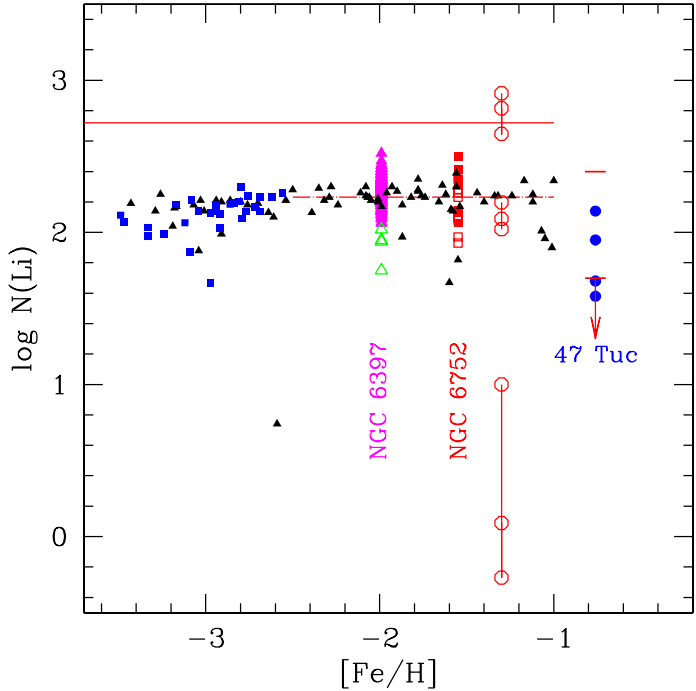} 
\caption{
 Lithium abundances as a function of [Fe/H] in halo stars and in scarcely 
 evolved stars in three GCs. Halo data are from \cite[Mel\'endez et al. (2009)]{melendez2009}, 
 represented as black triangles (non LTE models). (Blue) full squares are from 
 Sbordone et al. (2009), analyzed by 3D non LTE models. The top horizontal line 
 represent a WMAP -- standard Big Bang nucleosynthesis value $\log 
 \epsilon$(Li)=2.72, the dot--dashed line represents an eye fit of the \cite[Mel\'endez et al. (2009)]{melendez2009} data in the range of the GC metallicities. Data for NGC~6397 are from 
 \cite[Lind et al. (2009)]{lind2009}. The three open triangles are relative to 
 the data for three subgiants, and may not represent the turnoff abundances in 
 this cluster. Data for NGC~6752 are from Pasquini et al. (2005), plotted as open 
 or full squares according to the two different temperature scales used in 
 their work. The full circles are the data for 47 Tuc by Bonifacio et al. (2007). 
 The limits of the lithium range in the 50 stars recently examined by D'Orazi (these proceedings) are also given. Open circles at [Fe/H]=--1.3 represent the 
 average abundances in the ejecta of models of 4, 5 an 6\msun for three 
 different mass loss rate formulations (see text). }
   \label{f6}
\end{center}
\end{figure}

Figure \ref{f6} shows a compact summary of what we know about lithium 
abundances in the halo and in GCs in the plane $\log \epsilon$(Li) versus 
[Fe/H]. The halo stars are plotted as triangles, from \cite[Mel\'endez et al. (2009)]{melendez2009} (their non LTE abundances are plotted). The data for three clusters are added, 
at their [Fe/H] content, taken from \cite{carretta2009c} scale. The references 
for the clusters data are in the figure label. Notice that the three open 
triangles of NGC~6397, at much lower $\epsilon$(Li) than the other points, 
refer to subgiants, in which lithium can be reduced by mixing. Although the 
data analysis is not homogeneous among the different samples, the figure shows 
interesting trends. The lithium spread of the halo stars in the range of 
metallicities of the clusters NGC~6397 and NGC~6752 is very small around a 
plateau value $\log \epsilon$(Li)$\sim$2.2. In fact the full triangles at log 
$\epsilon$(Li)$<2$\ are lower mass stars for which depletion is expected 
(\cite[Mel\'endez et al. 2009]{melendez2009}). The WMAP -- big bang 
nucleosynthesis ``standard" abundance, $\log \epsilon$(Li)=2.72 (e.g. 
\cite[Cyburt et al. 2009]{cyburt2009}) is much larger than the plateau 
abundance.
The lithium spread in the clusters appears a bit larger, although 
\cite{lind2009} point out that in NGC~6397 it is consistent with the 
observational error. We should expect a larger lithium spread among GC stars if 
there are SG stars, even if the pollutors' gas (AGB or massive stars envelopes) 
has been diluted with pristine gas (\cite[Decressin et al. 
2007b]{decressin2007b}, \cite[Prantzos et al. 2007]{prantzos2007}). The 
dilution is very plausible if there is a direct correlation between lithium and 
sodium abundances, as convincingly shown in NGC~6752 (\cite[Pasquini et al. 
2005]{pasquini2005}). A similar correlation also appears in NGC~6397, but it is 
based only on the high sodium abundance of the three subgiants plotted as open 
triangles (\cite[Lind et al. 2009]{lind2009}). A possible anticorrelation among 
the stars of 47~Tuc (\cite[Bonifacio et al. 2007]{bonifacio2007}) is not 
convincing, as these stars may be subject to lithium depletion mechanisms due 
to their larger iron content (D'Orazi, these proceedings). In 
addition, according to \cite{pasquini2008}, two stars in NGC~6397 differ by 
$\sim$0.6dex in oxygen, but have ``normal" $\log \epsilon$(Li)$\sim$2.2: this 
is certainly not easily compatible with a simple dilution model, and may 
require that the pollutors are also important lithium producers. In fact, if 
the AGB pollutors produce enough lithium, a dilution model must take it into 
account.

Notice that the dilution model is not so straightforward as we may think a 
priori: it will include a fraction $\alpha$\ of matter with pristine Li, plus a 
fraction (1-$\alpha$) having the Li of the ejecta (so, either the abundance of 
the AGB ejecta in the AGB mass range involved in the SG formation, or zero Li 
for the FRMS model). The dilution required to explain a given range in observed 
Li is different if we assign to the pristine Li the value $\log 
\epsilon$(Li)=2.72 (see above), or the atmospheric Pop II value ($\sim$2.2), or 
some intermediate value. In addition, if we are assuming that the uniform 
surface abundance of Li in Pop II is due to a depletion mechanism, also the 
abundance resulting from the dilution model must be decreased to take into 
account a similar depletion factor.

If we take our ``standard mass loss" results of Fig. \ref{f6} at face value, 
ignoring the big question mark on mass loss, the yields can be used to predict 
the lithium expected in the SG, if the SG is a result of star formation from 
AGB ejecta diluted with pristine gas. The abundances will depend mainly on the 
mass range of the AGB progenitors: if the ejecta of masses in the range 4.5 --
6\msun\ are involved, their abundance is $\log \epsilon$(Li)$\sim$2, 0.7dex 
smaller than the Big Bang abundance. In order to explain the abundances 
observed in NGC~6397 or in NGC~6752, a dilution model including the ejecta of 
these AGB stars will require a percentage of pristine matter only {\sl slightly 
smaller} than in a model including the lithium free FRMS, and we will not be 
able to discriminate between the two models.  The case is different if the Big 
Bang abundance is ``non standard" and closer to the observed halo stars average 
value.

A different interesting problem is posed by the GCs in which a ``blue" main 
sequence (MS) has been revealed from precise HST photometry, namely \ocen\ 
(\cite[Bedin et al. 2004]{bedin2004}) and NGC~2808 (\cite[D'Antona et al. 
2005]{dantona2005}, \cite[Piotto et al. 2007]{piotto2007}). The blue MS can only 
be interpreted as a very high helium MS (mass fraction Y$\sim$0.38) 
(\cite[Norris 2004]{norris}, \cite[Piotto et al. 2005]{piotto2005}). Actually, 
in NGC~2808 three MS well separated each other in color are present 
(\cite[Piotto et al. 2007]{piotto2007}), corresponding to three main helium 
content values, and in agreement with the predictions made from the 
distribution of stars in the very extended and multimodal horizontal branch 
(see, e.g. \cite[D'Antona \& Caloi 2004]{dc2004}, \cite[D'Antona et al. 
2005]{dantona2005}). \cite{pumo2008} noticed that the helium abundances of 
super--AGB stars envelopes are within the small range 0.36$<$Y$<$0.38 
(\cite[Siess 2007]{siess2007}) and \cite[D'Ercole et al. (2008)]{dercole2008} 
have shown that a full chemo--hydrodynamical model of the cluster can provide a 
reasonable interpretation of the three MSs of NGC~2808, {\sl provided that the 
blue MS is formed directly by matter ejected from the super--AGB range, 
undiluted with pristine gas}. In the future, spectroscopic observations of the 
blue MS in \ocen\ and NGC~2808 will provide a falsification of this hypothesis, 
e.g. by means of the oxygen and sodium abundance revealed. In particular 
lithium can be an important test too, as it could provide an independent 
calibration of the mass loss rate in the super--AGB phase. Already some 
observations of the turnoff stars in \ocen\ are available (Bonifacio, in 
this book), but it is not clear whether stars belonging to the blue MS have 
been observed. The ``standard mass--loss" super--AGB models shown in Fig. 
\ref{f5} predict that lithium in these stars may become very large if some blue 
MS stars are formed from the ejecta of the upper mass range of super--AGB 
stars. We need observations of the blue MS to falsify this prediction.

\vskip 0.3cm 
\thanks{We thank Corinne Charbonnel and the organizing committees 
for the invitation and for  the successful and intense meeting. We are grateful 
to J. Mel\'endez and L. Sbordone for allowing us to use their data in advance of 
publication, and to V. D'Orazi and D. Romano for useful information.}

\end{document}